\documentclass[aps,prc,nofootinbib,twocolumn,superscriptaddress,showpacs,floatfix]{revtex4}
\usepackage{mathrsfs}
\usepackage{latexsym}
\usepackage{amsmath}
\usepackage{amssymb}
\usepackage{graphicx}
\usepackage{longtable}

\usepackage{bbm}
\usepackage{epsfig}
\usepackage[usenames]{color}
\usepackage[colorlinks=true,citecolor=blue,linkcolor=blue]{hyperref}

\begin{document}



\title{Isoscalar-vector interaction and {\bf hybrid} quark core in massive neutron stars}

\author{G. Y. Shao}
\affiliation{Department of Applied Physics, Xi'an Jiao Tong
University, Xi¡¯an 710049, China}

\author{M. Colonna}
\affiliation{INFN-Laboratori Nazionali del Sud, Via S. Sofia 62, I-95123
Catania, Italy}

\author{M. Di Toro}
\affiliation{INFN-Laboratori Nazionali del Sud, Via S. Sofia 62, I-95123
Catania, Italy}
\affiliation{Physics and Astronomy Dept., University of Catania, Via S. Sofia 64, I-95123
Catania, Italy}

\author{Y. X. Liu}
\affiliation{Department of Physics and State Key Laboratory of \\
Nuclear Physics and Technology,
Peking University, Beijing 100871, China}

\affiliation{Center of Theoretical Nuclear Physics,\\ National Laboratory of
Heavy Ion Accelerator, Lanzhou 730000, China}

\author{B. Liu}
\affiliation{IHEP, Chinese Academy of Sciences, Beijing, China}
\affiliation{Theoretical Physics Center for Scientific Facilities, \\Chinese
Academy of Sciences, Beijing, China}


\begin{abstract}
The hadron-quark phase transition in the core of massive neutron
stars is studied with a newly constructed two-phase model. For
nuclear matter, a nonlinear  Walecka type model with general
nucleon-meson and meson-meson couplings, recently calibrated by
Steiner, Hemper and Fischer, is taken. For quark matter, a modified
Polyakov-Nambu--Jona-Lasinio~(mPNJL) model, which gives consistent
results with lattice QCD data, is used. Most importantly, we
introduce an isoscalar-vector interaction in the description of
quark matter, and we study its influence on the hadron-quark phase
transition in the interior of massive neutron stars. With the
constraints of neutron star observations, our calculation shows that
the isoscalar-vector interaction between quarks is indispensable if
massive hybrids star exist in the universe, and its strength
determines the onset density of quark matter, as well as the
mass-radius relations of hybrid stars. Furthermore, as a connection
with heavy-ion-collision experiments we give some discussions about
the strength of isoscalar-vector interaction and its effect on the
signals of hadron-quark phase transition in heavy-ion collisions, in
the energy range of the NICA at JINR-Dubna and FAIR at GSI-Darmstadt
facilities.

\end{abstract}

\pacs{26.60.Kp,21.65.Qr, 97.60.Jd}

\maketitle

\section{Introduction}
The Equation of State~(EoS) of neutron star matter is closely
associated with particles that appear in neutron stars. Since the
matter in the core of neutron stars is possibly compressed to
several times of the saturation nuclear density, new particles and
even hadron-quark phase transitions may appear in the interior of
these compact objects \cite{Glendenning99,Menezes,Brown06,
Maruyama06,Shao09,Muto,Blaschke1999,Baldo03,
Weber2005,Alford20057,HShen081,Lattimer2007,Klahn2007}. On the other
hand, the EoS of neutron star matter is crucial for the macroscopic
features of neutron stars. Each EoS corresponds to one unique
mass-radius relation of neutron stars by solving the
Tolman-Oppenheimer-Volkoff (TOV) equation \cite{Tolman39}.

Interestingly, more and more neutron star observations, especially,
the accurate measurement of the pulsar J1614-2230 with the mass
$1.97\pm0.04\,\texttt{M}_\odot$ \cite{Demorest10}, the astrophysical
observations of X-ray bursts \cite{Guver10,Ozel11,Ozel09,Guver102}
and thermal emissions from quiescent low-mass X-ray binaries (LMXBs)
in the globular clusters \cite{Webb07,Heinke06,Guillot11}, gradually
provide a reliable constraint on the mass-radius relations which  is
tightly connected to the EoS of neutron star matter. The analysis of
these astrophysical observations shows that the radius of a 1.4
solar mass neutron star lies between 10.4 and 12.9 km, independent
of assumptions about the composition of the core \cite{Steiner12,
Steiner10}. The relatively small radius of  1.4 solar mass neutron
stars means the EoS near the suturation density is soft, and the
discovery of massive neutron star J1614-2230 requires the EoS is
stiff at high densities. The combination of these constraints rules
out many EoSs of hadron models. Besides, experimental information
from heavy-ion collisions (HIC)~\cite{Danielewicz02, Gupta11} and
lattice QCD simulations~\cite{Karsch01, Karsch02, Allton02,
Kaczmarek05, Cheng06, YAoki99, Borsanyi10} are also available to put
some constraints on the EoSs of nuclear and quark matter.

All these progresses on astrophysical observations and laboratory
nuclear experiments promote scientists to explore the relevant
physics behind. The hadron-quark phase transition is one of the most
concerned topics, and so far it is still controversial whether
quarks can appear in cold neutron star
\cite{Ozel10,Alford07,Fraga01,Ruster04,Orsarie13}. It is also an
important topic in heavy-ion collisions, and related experiments at
medium and high densities will be performed in the near future on
the updated facilities of  NICA at JINR-Dubna and FAIR at
GSI-Darmstadt.

The hybrid neutron star picture, with
 direct quark contributions in the inner core, seems to have problems in describing
 large mass neutron-stars. All that is due to a lack of repulsion at high baryon
 densities of the present quark matter effective interactions \cite{Burgio13}.
 This is in fact the main point which has motivated the present paper. 

 Hadron models based on the
relativistic mean-field~(RMF) theory are usually used to study the
properties of finite nuclei and nuclear matter. But in literature
with the same parameter sets, the obtained mass-radius ($M-R$)
relations with or without a hadron-quark phase
transition~(e.g.,~\cite{Steiner00,Dexheimer10,Shao110,Mahajan84,HShen081})
are not supported by the recent analysis of neutron star
observations~\cite{Steiner12, Steiner10}. To solve the problem of
inconsistency between neutron star observations and nuclear matter
EoS in RMF theory, in this study we will investigate the
hadron-quark phase transition with a newly constructed two-phase
model. To describe nuclear matter, we take the extended Walecka
model with general nucleon-meson and  meson-meson couplings recently
calibrated by Steiner, Hemper and Fischer \cite{Steiner122}, which
describes well the properties of nuclei and nuclear matter, and
supports recent simulations of the supernovae dynamics. For quark
matter, we take the mPNJL quark
model~\cite{Dexheimer10,Blaschke10,Shao110} which shares with QCD
global symmetries and the phenomenon of chiral symmetry breaking as
well as  an effective (de)confinement at finite densities and
temperatures.

Most importantly, in the description of quark matter, we focus on
the inclusion of the isoscalar-vector interaction and its influence
on the hadron-quark phase transition in the interior of massive
neutron stars. With the constraints of neutron star observations,
our calculation shows that the isoscalar-vector channel
interaction is needed if massive hybrid stars exist in the universe,
and its strength determines the onset density of quark matter, as
well as the mass-radius relation of hybrid stars. Our previous
study~\cite{Shao12} also illustrates that this channel interaction
affects the hadron-quark phase transition in heavy-ion collisions at
finite temperatures and moderate densities. Therefore, as a
connection with heavy-ion-collision experiments, we further give
some discussions about the strength of isoscalar-vector interaction
and its influence on the possible phase transition signals from
asymmetric nuclear matter to quark matter in heavy-ion collision
experiments.

The paper is organized as follows. In Section II, we describe
briefly the two-phase model and give the relevant formulas of the extended nonlinear Walecka model
and the mPNJL model. In Section III, we present the numerical results, and
give some discussions about the phase transition in massive neutron stars,
as well as the connection with the phase transition in heavy-ion collisions.
Finally, a summary is given in Section IV.

\section{ The models}
In the two-phase model, the pure hadronic phase and quark phase are
described by the nonlinear Walecka type model and the mPNJL model,
respectively. As for the coexisted phase between the pure hadronic
phase and quark phase, the two phases are connected through
the Gibbs conditions with the thermal, chemical and mechanical
equilibriums, as well as the global charge neutrality
condition~\cite{Glendenning91}.
\subsection{The hadronic model}
Recently one new equation of state of nuclear matter, labeled SFHO,
was constructed based on the extended non-linear Walecka model in
RMF theory, and it was taken to simulate core-collapse
supernova~\cite{Steiner122}. The obtained results satisfy the
requirements of nuclear physics and match well the astrophysical
observations. The Lagrangian of this model is written as
\begin{widetext}
\begin{eqnarray}
\cal{L}^H_{} &=&\sum_N\bar{\psi}_N[i\gamma_{\mu}\partial^{\mu}- M
          +g_{\sigma }\sigma
          -g_{\omega }\gamma_{\mu}\omega^{\mu}
          -g_{\rho }\gamma_{\mu}\boldsymbol\tau_{}\cdot\boldsymbol
\rho^{\mu}]\psi_N  +\frac{1}{2}\left(\partial_{\mu}\sigma\partial^
{\mu}\sigma-m_{\sigma}^{2}\sigma^{2}\right)
           \nonumber\\
       & &{}- V(\sigma)+\frac{1}{2}m^{2}_{\omega} \omega_{\mu}\omega^{\mu}
          -\frac{1}{4}\omega_{\mu\nu}\omega^{\mu\nu}
          +\frac{1}{2}m^{2}_{\rho}\boldsymbol\rho_{\mu}\cdot\boldsymbol
\rho^{\mu}
          -\frac{1}{4}\boldsymbol\rho_{\mu\nu}\cdot\boldsymbol\rho^{\mu\nu}\nonumber \\
      & &{}+\frac{\zeta}{24}  g_\omega^4(\omega^\mu \omega_\mu)^2+\frac{\xi}{24}
       g_\rho^4(\boldsymbol\rho^\mu \cdot \boldsymbol\rho_\mu)^2+g_\rho^2 f(\sigma,
       \omega^\mu \omega_\mu)\boldsymbol\rho^\mu \cdot \boldsymbol\rho_\mu
      \nonumber\\
      & &{}+\sum_l\bar{\psi_l}(i\gamma_{\mu}\partial^{\mu}- m_l)\psi_l    ,
 \end{eqnarray}
\end{widetext}
where $
\omega_{\mu\nu}= \partial_\mu \omega_\nu - \partial_\nu
\omega_\mu$, $ \rho_{\mu\nu} \equiv\partial_\mu
\boldsymbol\rho_\nu -\partial_\nu \boldsymbol\rho_\mu$, and the scalar meson potential
$V(\sigma)= \frac{1}{3} b\,(g_{\sigma} \sigma)^3+\frac{1}{4} c\,
(g_{\sigma} \sigma)^4$. $f(\sigma, \omega^\mu \omega_\mu$) takes the general form as
\begin{equation}
f=\sum_{i=1}^6 a_i\sigma^i+\sum_{j=1}^3 b_j(\omega_\mu \omega^\mu)^j.
\end{equation}
It was first introduced in \cite{Steiner06} to provide additional
freedom in varying the symmetry energy, and the parameters have been
recalibrated recently to fulfill the constraints of nuclear physics
and astrophysical observations. 
Due
to the uncertainties of hyperon-meson couplings, only protons and
neutrons are considered in this study. Such point will be further
discussed in Sect.III. Electrons and muons, last term of Eq.(1), are
included in the calculation in order to keep the charge neutrality
of neutron star matter.

Under the mean field approximation,
the meson field equations  can be obtained as
\begin{equation}\label{sigma}
m_\sigma^2\sigma-g_\sigma\rho_S^{}+bmg_\sigma^3\sigma^2
+cg_\sigma^4\sigma^3-g_\rho^2\rho^2\frac{\partial f}{\partial \sigma}=0
\end{equation}
\begin{equation}\label{omega}
m_\omega^2\omega-g_\omega\rho_B^{}+\frac{\zeta}{6}g_\omega^4\omega^3
+g_\rho^2\rho^3\frac{\partial f}{\partial \omega}=0
\end{equation}
\begin{equation}
m_\rho^2\rho+\frac{1}{2}g_\rho\sum\tau_{3i}^{}\rho_i^{}+2g_\rho^2\rho f
+\frac{\xi}{6}g_\rho^4\rho^3=0
\end{equation}

In Eqs. (\ref{sigma}-\ref{omega}),
\begin{equation}
\rho_S^{}=\sum_{i=p,n}2\int\frac{d^3k}{(2\pi)^3}\frac{M^*}{E^*}[f_i(k)-\bar{f}_i(k)]
\end{equation}
\begin{equation}
\rho_B^{}=\sum_{i=p,n}2\int\frac{d^3k}{(2\pi)^3}[f_i(k)-\bar{f}_i(k)]
\end{equation}
are the scalar and baryon densities, respectively.

The thermodynamical potential of the nucleon-meson system is
\begin{widetext}
\begin{eqnarray}
\Omega^H_{} &=&-\beta^{-1}\sum_{i=N,l} 2 \int \frac{d^3 k}{(2\pi)^3}\bigg{\{}\texttt{ln}\bigg[1+e^{-\beta(E_i^\star (k)-\mu_i^\star)}\bigg]
+\texttt{ln}\bigg[1+e^{-\beta(E_i^\star (k)+\mu_i^\star)}\bigg{]}\bigg{\}}+\frac{1}{2}m_\sigma^2 \sigma^2-\frac{1}{2}m_\omega^2 \omega ^2 -\frac{1}{2}m_{\rho}^2\rho_{}^2
 \nonumber\\
 & &{} +V(\sigma)
    -\frac{\zeta}{24}  g_\omega^4\omega^4-\frac{\xi}{24}  g_\rho^4\rho^4-g_\rho^2 \rho^2f
          ,
 \end{eqnarray}
\end{widetext}
where $E_i^{*}(k)=\sqrt{k^2+(M_i-g_\sigma \sigma)^2}$, $\mu_{i}^{*}=\mu_{i}-g_{\omega
}\omega-g_{\rho}\tau_{3i}^{}\rho^{} $ for nucleons;
$E_i^{*}(k)=\sqrt{k^2+m_i^2}$, $\mu_{i}^{*}=\mu_{i}$ for leptons.
The corresponding energy density and pressure of nuclear matter can be
derived as
%
%
\begin{eqnarray}
\varepsilon^{H} &= & \sum_{i=N,l}\frac{2}{(2\pi)^3} \int \! d^3
 k E_i^{*}(k)[f_{i}(k)+\bar{f}_{i}(k)]
+g_\omega\omega\rho_B^{^{}} \nonumber\\
& &{}+\frac{1}{2}g_\rho\rho(\rho_p-\rho_n)
+\frac{1}{2}m_\sigma^2
\sigma^2 +V(\sigma)\nonumber \\
& &{}-\frac{1}{2}m_\omega^2 \omega^2 -
\frac{1}{2}m_\rho^2 \rho^2 -\frac{\zeta}{24}  g_\omega^4\omega^4 \nonumber \\
& &{}-\frac{\xi}{24}  g_\rho^4\rho^4+g_\rho^2 \rho^2f \,
 \, ,
\end{eqnarray}
\begin{eqnarray}
P^{H} & = & \sum_{i=N,l} \frac{1}{3}\frac{2}{(2\pi)^3} \int \!
d^3  k \frac{k^2}{E_i^{*}(k)}[f_{i}(k)+
\bar{f}_{i}(k)]\,\,\,\,\,\,\,\,\,\, \,\,\,\,\,\,\,\,\,\, \nonumber \\
& &{}- \frac{1}{2}m_\sigma^2
\sigma^2 -V(\sigma)
+\frac{1}{2}m_\omega^2 \omega^2 +
\frac{1}{2}m_\rho^2 \rho^2 \nonumber \\
& &{}+\frac{\zeta}{24}  g_\omega^4\omega^4+\frac{\xi}{24}  g_\rho^4\rho^4+g_\rho^2 \rho^2f \, ,
\end{eqnarray}
%
where $f_{i}(k)$ and $\bar{f}_{i}(k)$ are the fermion and antifermion
distribution functions:
\begin{equation}
  f_{i}(k)=\frac{1}{1+\texttt{exp}[(E_{i}^{*}(k)-\mu_{i}^{*})/T]} ,
\end{equation}
\begin{equation}
  \bar{f}_{i}(k)=\frac{1}{1+\texttt{exp}[(E_{i}^{*}(k)+\mu_{i}^{*})/T]}.
\end{equation}
The model parameter set labeled SFHO
is used in the calculation, which can be fixed by
fitting the properties of symmetric nuclear matter at saturation
nuclear density. For the details of model parameters, on can refer \cite{Steiner122}

\subsection{The quark model}

For the quark phase, we take the modified three-flavor PNJL model with
the Lagrangian
\begin{eqnarray}
\mathcal{L}^{Q}&=&\bar{q}(i\gamma^{\mu}D_{\mu}-\hat{m}_{0})q+
G\sum_{k=0}^{8}\bigg[(\bar{q}\lambda_{k}q)^{2}+
(\bar{q}i\gamma_{5}\lambda_{k}q)^{2}\bigg]\nonumber\\
           &&-K\bigg[\texttt{det}_{f}(\bar{q}(1+\gamma_{5})q)+\texttt{det}_{f}
(\bar{q}(1-\gamma_{5})q)\bigg]\nonumber \\ \nonumber \\
&&-\mathcal{U}(\Phi[A],\bar{\Phi}[A],T)-G_V(\bar{q}\gamma^\mu q)^{2},
\end{eqnarray}
where $q$ denotes the quark fields with three flavors, $u,\ d$, and
$s$, and three colors; $\hat{m}_{0}=\texttt{diag}(m_{u},\ m_{d},\
m_{s})$ in flavor space; $G$ and $K$ are the four-point and
six-point interacting constants, respectively. The isoscalar-vector
interaction channel is also included.
As shown by Eq.(13), the $\mu = 0$ component of the isoscalar vector interaction corresponds to the density
operator $(\bar{q}\gamma^0 q)^{2}$, therefore it is conceivable to expect that the
finite-density environment brings a significant contribution to this channel.
Moreover, it is also well known that
standard (P)NJL models lead to a chiral restoration transition at unphysical
baryon densities just above the saturation point  \cite{Bub2005},  
easily reached in heavy ion  collisions, without any evidence of such effect. The presence of a repulsive vector
field, as considered in the present work, moves the transition to more realistic, higher densities.

This channel interaction
reduces the effective quark chemical potential,
$\tilde{\mu}=\mu-2G_Vn_q$. In this study we will focus on its
influence on the hadron-quark phase transition in dense neutron star
matter.

The covariant derivative in the Lagrangian is defined as $D_\mu=\partial_\mu-iA_\mu$.
The gluon background field $A_\mu=\delta_\mu^0A_0$ is supposed to be homogeneous
and static, with  $A_0=g\mathcal{A}_0^\alpha \frac{\lambda^\alpha}{2}$, where
$\frac{\lambda^\alpha}{2}$ is $SU(3)$ color generators.
The effective potential $\mathcal{U}(\Phi[A],\bar{\Phi}[A],T)$ is expressed in terms of the traced Polyakov loop
$\Phi=(\mathrm{Tr}_c L)/N_C$ and its conjugate
$\bar{\Phi}=(\mathrm{Tr}_c L^\dag)/N_C$. The Polyakov loop $L$  is a matrix in color space
\begin{equation}
   L(\vec{x})=\mathcal{P} \mathrm{exp}\bigg[i\int_0^\beta d\tau A_4 (\vec{x},\tau)   \bigg],
\end{equation}
where $\beta=1/T$ is the inverse of temperature and $A_4=iA_0$.

Different effective potentials were
adopted in literatures~\cite{Ratti06,Robner07,Fukushima08,Dexheimer10}.
The modified chemical dependent one
\begin{eqnarray}
    \mathcal{U}&=&(a_0T^4+a_1\mu^4+a_2T^2\mu^2)\Phi^2\nonumber \\
               & &a_3T_0^4\,\mathrm{ln}(1-6\Phi^2+8\Phi^3-3\Phi^4)
\end{eqnarray}
was used in~\cite{Dexheimer10, Blaschke10} which is a simplification of
\begin{eqnarray}\label{U}
    \mathcal{U}&=&(a_0T^4+a_1\mu^4+a_2T^2\mu^2)\bar{\Phi}\Phi\nonumber \\
               & &a_3T_0^4\,\mathrm{ln}\bigg[1-6\bar{\Phi}\Phi+4(\bar{\Phi}^3+\Phi^3)-3(\bar{\Phi}\Phi)^2\bigg]
\end{eqnarray}
because the difference between $\bar{\Phi}$ and $\Phi$ is smaller
at finite chemical potential, and $\bar{\Phi}=\Phi$ at $\mu=0$. In the calculation we will take
the form given in  equation.~\ref{U}, as used in \cite{Shao110}.
The related parameters, $a_0=-1.85,\,a_1=-1.44\times10^{-3},\,a_2=-0.08,\,a_3=-0.4$,
are still  taken from~\cite{Dexheimer10},
which can reproduce well the data obtained in lattice QCD calculation.

In the mean-field approximation, quarks can be taken as free quasiparticles
with constituent masses $M_i$,
and the dynamical quark masses~(gap equations) are obtained as
\begin{equation}
M_{i}=m_{i}-4G\phi_i+2K\phi_j\phi_k\ \ \ \ \ \ (i\neq j\neq k),
\label{mass}
\end{equation}
where $\phi_i$ stands for quark condensate.

The thermodynamic potential of quark matter in the mean-field level can be derived as
\begin{widetext}
\begin{eqnarray}
\Omega^{Q}_{}&=&\mathcal{U}(\bar{\Phi}, \Phi, T)+2G\left({\phi_{u}}^{2}
+{\phi_{d}}^{2}+{\phi_{s}}^{2}\right)-G_V(n_u+n_d+n_s)^2-4K\phi_{u}\,\phi_{d}\,\phi_{s}-2\int_\Lambda \frac{\mathrm{d}^{3}p}{(2\pi)^{3}}3(E_u+E_d+E_s) \nonumber \\
&&-2T \sum_{u,d,s}\int \frac{\mathrm{d}^{3}p}{(2\pi)^{3}} \mathrm{ln}\,\bigg[A(\bar{\Phi},\Phi,E_i-\mu_i,T)\bigg]-2T \sum_{u,d,s}\int \frac{\mathrm{d}^{3}p}{(2\pi)^{3}} \mathrm{ln}\,\bigg[\bar{A}(\bar{\Phi},\Phi,E_i+\mu_i,T)\bigg],
\end{eqnarray}
\end{widetext}
where $A(\bar{\Phi},\Phi,E_i-\mu_i,T)=1+3\Phi e^{-(E_i-\mu_i)/T}+3\bar{\Phi} e^{-2(E_i-\mu_i)/T}+e^{-3(E_i-\mu_i)/T}$
and $\bar{A}(\bar{\Phi},\Phi,E_i+\mu_i,T)=1+3\bar{\Phi} e^{-(E_i+\mu_i)/T}+3\Phi e^{-2(E_i+\mu_i)/T}+e^{-3(E_i+\mu_i)/T}$.

The values of $\phi_u, \phi_d, \phi_s, \Phi$ and $\bar{\Phi}$ are determined by minimizing the thermodynamical
potential
\begin{equation}
\frac{\partial\Omega^Q}{ \partial\phi_u}=\frac{\partial\Omega^Q}{\partial \phi_d}=\frac{\partial\Omega^Q}{\partial \phi_s}=\frac{\partial\Omega^Q}{\partial \Phi}=\frac{\partial\Omega^Q}{\partial \bar\Phi}=0.
\end{equation}
All the thermodynamic quantities relevant to the bulk properties of quark matter can be obtained from $\Omega_Q$. Particularly, the pressure
can be derived with $P=-(\Omega^Q(T,\mu)-\Omega^Q(0,0))$.
From Eq.(18) it is clear that
the introduction of the isoscalar vector interaction in the PNJL model
may give important contributions to the
pressure of quark matter. Indeed the EOS may become  much stiffer at high
densities for large $G_V$ values. A similar behavior can be obtained in the MIT bag
model if such term is included.

As an effective model, the (P)NJL model is not
renormalizable, so a cut-off $\Lambda$ is implemented in 3-momentum
space for divergent integrations. The model parameters:
$\Lambda=603.2$ MeV, $G\Lambda^{2}=1.835$, $K\Lambda^{5}=12.36$,
$m_{u,d}=5.5$  and $m_{s}=140.7$ MeV, determined
by fitting $f_{\pi},\ M_{\pi},\ m_{K}$ and $\ m_{\eta}$ to their
experimental values~\cite{Rehberg95}, are used in the calculation.

Concerning the strength of the isoscalar-vector coupling $G_V$,
there are no explicit constraints at finite density.
Some efforts to estimate a possible range of values for
this coupling
are briefly described below.
For the convenience to compare $G_V$ with
the strength of the isoscalar-scalar interaction $G$, and for later discussion, we define $R_V=G_V/G$.

A very naive estimation, based on the value taken by this ratio in the hadronic sector, would give
$R_V$ around 0.6-0.7 \cite{BoLiu2011}.
In Ref.\cite{Shao12}
a vector/scalar coupling ratio around $R_V=0.2$ was obtained from an
evaluation of only the Fock contributions of the scalar channels.
Values in the range $0.25<R_V<0.5$ are derived
by a Fierz transformation of an effective one-gluon exchange interaction, with $G_V$
depending on the strength of the $U_A(1)$ anomaly in the two-flavor model \cite{Klevansky92,Kashiwa12}.
However,  the point
is  that the coupling strength of the direct term cannot be fixed, so the
total effect of the isoscalar vector interaction is still unknown.
Other attempts to estimate $G_V$ are based on the fit of the vector meson spectrum~\cite{Vogl91}.
However,
the relation between the vector coupling in dense quark matter and the meson spectrum
in vacuum is expected to be strongly modified by in-medium effects \cite{Fukushima08,ZZhang09}.

Because of the uncertainties discussed above,
we will treat $R_V$ as a free parameter.
Aim of the present work is just to get
some relevant information from neutron star properties. 
We notice that suggestions to catch
 information on the isoscalar vector interaction may also come from heavy ion reactions at
next-generation colliders,
such as NICA and FAIR \cite{Shao12}.
A combined study of the two aspects appears as a promising tool to get hints on
the strength of $R_V$.

\subsection{The hadron-quark phase {\bf coexistence}}
The Gibbs criteria are usually implemented for the phase
equilibrium of a complex system with more than one
conservation charge. The Gibbs conditions for the phase
coexistence within a hadron-quark transition in compact star are
\begin{eqnarray}
& &\mu_\alpha^H=\mu_\alpha^Q,\ \ \ \ \  \ T^H=T^Q,\ \ \ \ \ \ P^H=P^Q,
\end{eqnarray}
where $\mu_\alpha$ are usually chosen with $\mu_n$ and $\mu_e$.
Under the $\beta$ equilibrium without trapped neutrino, the chemical potential
of other particles including all baryons, quarks, and leptons can be derived by
\begin{equation}
    \mu_i=g_i\mu_n-q_i\mu_e,
\end{equation}
where $g_i$ and $q_i$ are the baryon number and electric charge
number of particle species $i$, respectively.

The baryon number density and energy density in the mixed phase are
composed of two parts with the following combinations
\begin{equation}
    \rho=(1-\chi)\rho_B^H+\chi\rho_B^Q,
\end{equation}
and
\begin{equation}
    \varepsilon=(1-\chi)\varepsilon^H+\chi\varepsilon^Q,
\end{equation}
where $\chi$ is the volume fraction of quark matter. For the coexisted phase,
the electric neutrality is fulfilled globally
with
\begin{equation}
    q_{total}=(1-\chi)\sum_{i=B,l}q^i \rho_i+\chi \sum_{i=q,l}q^i \rho_i=0.
\end{equation}

\section{Numerical results and discussions}
\begin{figure}[htbp]
\begin{center}
\includegraphics[scale=0.26]{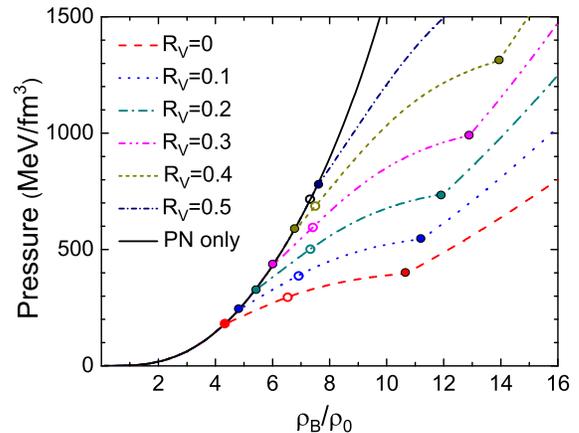}
\caption{\label{fig:EoS1}(color on line) EoSs of neutron star matter
without and with a hadron-quark phase transition for different
isoscalar-vector interaction coupling $G_V$. For each value of
$R_V$, the two solid
 dots with the same color indicate the range of the mixed phase, and the cycle marks
  the largest pressure that can be reached in the core of neutron star.}
\end{center}
\end{figure}

We present in Fig.~\ref{fig:EoS1} the EoSs of neutron star matter
without and with the hadron-quark phase transition for different
strength of the isoscalar-vector interaction. For each value of $R_V
= G_V/G$, the two solid dots with the same color indicate the range
of the coexisted phase, and the cycle marks the largest pressure
that can be reached in the core of neutron star by solving the TOV
equation.  With increasing vector strength in the quark sector the
onset of the transition is moving to higher densities since the
quark pressure is also increasing.

This figure demonstrates that the isoscalar-vector interaction of
quark matter plays an important role in the hadron-quark phase
transition. In particular the value of $R_V$ is crucial for the EoS
of neutron star matter at high densities. Besides, it shows that the
mixed range can be reached only inside massive neutron stars.  In
fact, for the case $R_V$ larger than 0.484, the calculation shows
that quark matter does not appear in the neutron star core.

\begin{figure}[htbp]
\begin{center}
\includegraphics[scale=0.26]{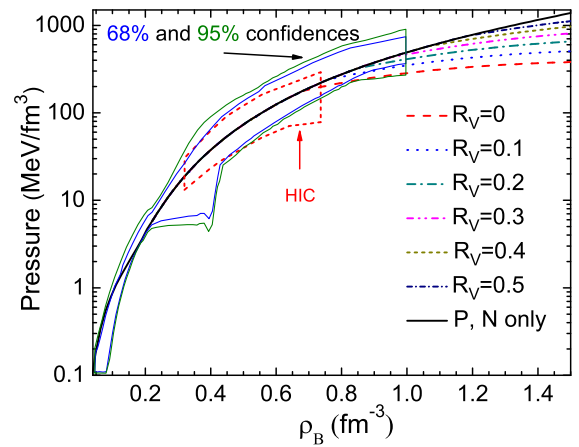}
\caption{\label{fig:EoS2}(color on line) The rescaled EoSs as in
Fig. \ref{fig:EoS1} with the constraints from heavy-ion collisions
and neutron star observations. The region labeled ¡°68\%¡±
(¡°95\%¡±) gives the 68\% (¡°95\%¡±) confidence level from eight
neutron star observation~(Ref.~\cite{Steiner12} and refs. therein).
 ¡°HIC¡± marks the constraints inferred from heavy-ion collisions~\cite{Danielewicz02}.
}
\end{center}
\end{figure}

To compare with the data from heavy-ion-collision experiments and
neutron star observations, we re-scaled in Fig. \ref{fig:EoS2} the
EoSs displayed in Fig. \ref{fig:EoS1}. It shows all the EoSs can
fulfill these constraints at low and moderate densities. However,
for the case of hybrid star at high densities, $R_V> 0.1$ appears to
be favored by recent neutron star observations of the mass-radius
correlation in $1\sigma$ contours. In the following we will
constrain the value of $R_V$ using the $M-R$ relations and the
accurately measured mass of Pulsar J1614-2230.

As already stated before, in the
present study hyperons are not included in the hadron sector mostly
because of the uncertainties on the hyperon-meson couplings in the
nuclear medium. When hyperons appear in the nuclear matter, just
from a degree of freedom counting we can expect a sudden softening
of the nuclear EoS. Such effect can be largely modified from the
hyperon interactions. Indeed this seems to be the indication
emerging from Heavy Ion data of Fig. \ref{fig:EoS2}: the hadronic
EoS appears to remain rather stiff at densities between 3 and 5
$\rho_0$, where hyperons would appear in the medium just from
chemical potential arguments. However we must also note that the
high density matter formed in HIC will certainly differ from the one
of neutron-stars, in particular for the lack of chemical
equilibrium.
In any case, mass-radius calculations for hybrid neutron stars seem
to be more sensitive to the model adopted for the quark phase than
to the hadronic EOS \cite{Burgio13,Baldo12}.

As a further theoretical study, to investigate the properties of  massive neutron stars with both hyperons
and quarks, one can fine-tune the meson-hyperon couplings and/or
introduce the strange meson mediated interaction between hyperons in the hadron sector, to
adjust the stiffness of the hadronic EOS.
This will be the object of a forthcoming analysis.



\begin{figure}[htbp]
\begin{center}
\includegraphics[scale=0.27]{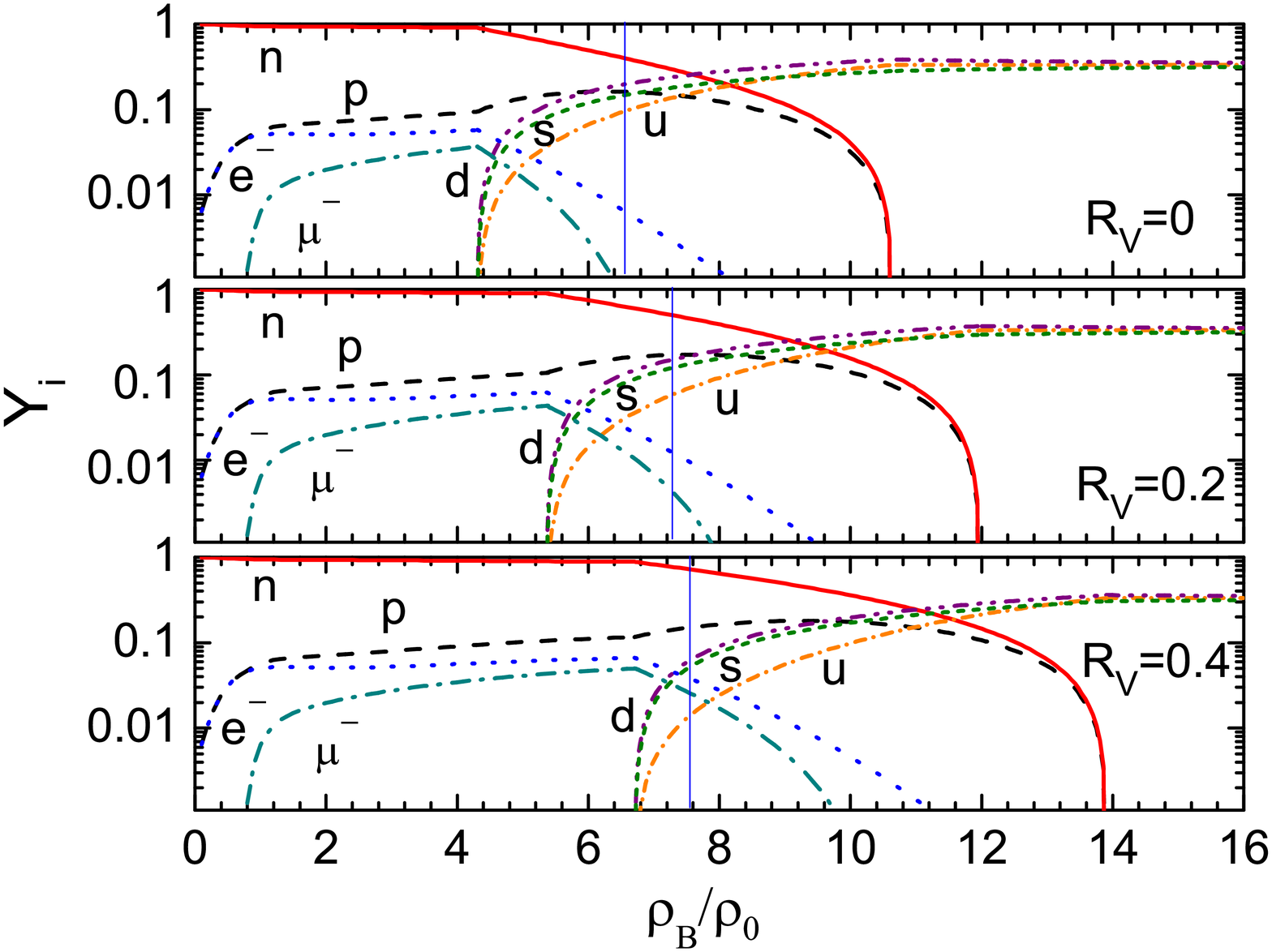}
\caption{\label{fig:Richness}(color on line) Relative fractions of different species as
functions of baryon density with $R_V=0,\,0.2,\,0.4$, respectively.
The blue vertical lines mark the corresponding central densities of hybrid stars.
}
\end{center}
\end{figure}

In order to see how the isoscalar-vector interaction affects the
threshold of the hadron-quark phase transition, we display in Fig.
~\ref{fig:Richness} the relative fractions of different species as
functions of baryon density for $R_V=0,\,0.2,\,0.4$, respectively.
This figure shows that a stronger isoscalar-vector interaction
postpones the onset density of quark matter, and the central density
of the corresponding hybrid star moves to a higher value. However a
larger $R_V$ also means that the fraction of quark matter is smaller
in the core of neutron star. Particularly, if $R_V$ is large enough,
the onset density of quark matter, in fact in the mixed phase, will
be larger than the central density of the neutron star. In this
case, no quarks can appear in the core of neutron stars. This
clearly demonstrates the crucial role that the isoscalar-vector
interaction plays on the hadron-quark phase transition in massive
neutron stars.

\begin{figure}[htbp]
\begin{center}
\includegraphics[scale=0.27]{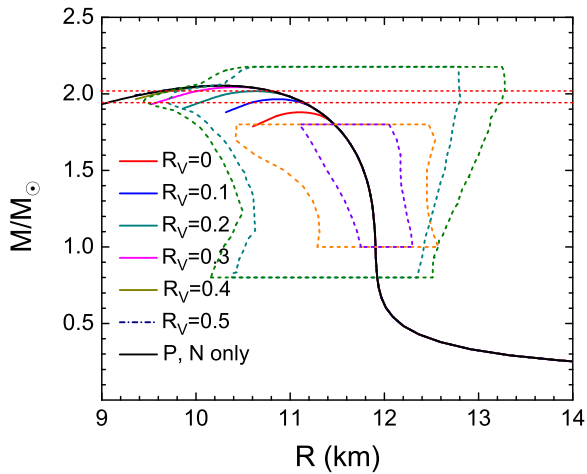}
\caption{\label{fig:R-M-NSO}(color on line) Mass-radius relations of neutron stars.
The solid curve is the result without quarks and the dash curves are
the results with a hadron-quark phase transition for different
$R_V$. The inner (outer) two contours  show the 1$\sigma$ and
2$\sigma$ confidence ranges of the $M-R$ relations  given
in~Ref.~\cite{Steiner10}~(Ref.~\cite{Steiner12}), based on six
(eight) neutron star observations of the X-ray bursts
\cite{Guver10,Ozel11,Ozel09,Guver102} and thermal emissions from
quiescent low-mass X-ray binaries (LMXBs) in the globular clusters
\cite{Webb07,Heinke06,Guillot11}.}
\end{center}
\end{figure}

In Fig.~\ref{fig:R-M-NSO} we plot the mass-radius relations of
hybrid stars with different $R_V$. The inner (outer) two contours
represent the 1$\sigma$ and 2$\sigma$ confidence ranges of the
$M-R$ relations  given
in~Ref.~\cite{Steiner10}~(Ref.~\cite{Steiner12}), based on six
(eight) neutron star observations of the X-ray bursts
\cite{Guver10,Ozel11,Ozel09,Guver102}
 and thermal emissions from quiescent low-mass X-ray binaries (LMXBs) in
the globular clusters \cite{Webb07,Heinke06,Guillot11}. This figure
shows that the EoS of neutron star matter with the parameter set of
SFHO fulfills well the constraints of neutron star observations.
This figure also gives us an explicit picture of how the strength of
the isoscalar-vector interaction influences the macroscopical
properties of massive hybrid stars. The radio timing observations of
the binary millisecond pulsar J1614-2230 with a strong general
relativistic Shapiro delay signature, implies that the pulsar mass
is $1.97\pm0.04\,\mathrm{M}_\odot$~\cite{Demorest10}. The discovery
of this massive pulsar rules out many soft EoSs. If the
isoscalar-vector interaction of quark matter is not included, the
maximum mass of hybrid stars is $1.88\,\mathrm{M}_\odot$, less than
the known maximum neutron star mass. However, with the inclusion of
this channel interaction and taking $R_V=0.055$, the obtained
maximum mass of a hybrid star can reach the lower mass limit of the
pulsar J1614-2230. Therefore, $R_V\geq 0.055$ is required for the
existence of massive hybrid stars in the universe, and no quarks
appear for $R_V> 0.484$ as shown in Fig.~\ref{fig:EoS1}.

Finally, we like to present some further discussions about the
isoscalar-vector interaction in quark matter. Apart the
general argument about the relevance of this term in hadronic matter
at low temperature and increasing baryon density, we have some
specific points from results of effective non-perturbative QCD
models.

When including this channel interaction in quark model, the value of
$R_V$ affects the location and emergence of critical points of
chiral symmetry
restoration~\cite{Fu08,Costa10,Carignano10,Lourencol12,Kashiwa12}. We
note that in all the present effective quark models with chiral
restoration, like NJL or PNJL, show, at zero temperature, the chiral
transition takes place at unphysical low baryon densities, just above the
saturation value. The inclusion of the isoscalar-vector term, which
increases pressure and kinetic chemical potential, will shift the
transition to more realistic density regions.

Moreover  this channel in the effective quark Lagrangian also
influences the onset densities and the expected
phase-transition signals from charge asymmetric nuclear matter to
quark matter in the two-phase model related to experiments in
heavy-ion collisions~\cite{Shao12}. New data about properties
of a mixed phase eventually probed in high density regions will
provide important information on the strength of the
isoscalar-vector term in the quark interaction.

In conclusion although this coupling presently cannot be
determined from experiments and  lattice QCD simulations, there are some
good hints about its existence:
\begin{itemize}
\item{Compared with the hadron Walecka model, the isoscalar-vector
interaction of quark matter plays a repulsive role similar to the
$\omega$ meson, important for the properties of nuclear matter in
Quantum Hadron Dynamics model, in particular at finite densities and
low temperatures.}
\item{This channel interaction can be derived
from higher order Fock (exchange) terms or Fierz transformations or
fitting the vector meson spectrum~\cite{Shao12}.}
\item{ The
existence of massive hybrid neutron stars as shown in this study.}
\end{itemize}

\section{summary}
We have studied the hadron-quark phase transition in dense neutron
star matter with an improved two-phase model. The calculations show
that massive hybrid stars possibly exist in the universe.
In this respect the isocalar-vector interaction between quarks is
crucial for the hadron-quark phase transition. Its strength
determines whether quarks can appear in the interior of neutron
stars.

Although the accurate value of $R_V$ is still not known by far,
neutron star observations can gradually provide some constraints on
it. In our previous study about the hadron-quark phase transition in
heavy-ion collisions, we have demonstrated that the inclusion of
this channel interaction postpones the onset density of quark
matter. The corresponding phase-transition signals, in particular on
properties of the mixed phase, in the case of charge asymmetric
nuclear matter to quark matter will be strengthened. Then, based on
the phase transition features of asymmetric strongly interacting
matter, we also proposed some suggestions to probe the phase
transition signals in relevant experiments at the FAIR and NICA
facilities~\cite{Shao12}.
The promising conclusion is that in a near future the
combination of neutron star observations and the energy scan of the
phase-transition signals at FAIR/NICA may provide us some
hints on the value of $R_V$, which is helpful for the understanding
of quark matter interactions and neutron star structure.

\begin{acknowledgments}
This  work  is supported by the National Natural Science Foundation
of China under Grants No. 11147144 and 11075037.
\end{acknowledgments}%

\end{document}